\documentclass[longauth]{aa}
\usepackage{graphicx}
\usepackage{txfonts}
\usepackage{hyperref}
\usepackage{natbib}
\usepackage{makecell}
\usepackage{lipsum}
\usepackage{subcaption}
\usepackage{lscape}  
\usepackage{placeins}

\newcommand{\affOAUW}{Astronomical Observatory, University of Warsaw, Al. Ujazdowskie 4, 00-478 Warszawa, Poland}
\newcommand{\affWARWICK}{Department of Physics, University of Warwick, Coventry CV4 7AL, UK}
\newcommand{\affWEIZMANN}{Department of Particle Physics and Astrophysics, Weizmann Institute of Science, Rehovot 76100, Israel}
\newcommand{\affVillanova}{Villanova University, Department of Astrophysics and Planetary Sciences, 800 Lancaster Ave., Villanova, PA 19085, USA}
\newcommand{\affRLCO}{Las Cumbres Observatory Global Telescope Network, 6740 Cortona Drive, suite 102, Goleta, CA 93117, USA}
\newcommand{\affRMill}{Millennium Institute of Astrophysics MAS, Nuncio Monsenor Sotero Sanz 100, Of. 104, Providencia, Santiago, Chile}
\newcommand{\affRSUPA}{SUPA, School of Physics \& Astronomy, University of St Andrews, North Haugh, St Andrews KY16 9SS, UK}
\newcommand{\affRLiver}{Astrophysics Research Institute, Liverpool John Moores University, Liverpool CH41 1LD, UK}
\newcommand{\affRIPAC}{IPAC, Mail Code 100-22, Caltech, 1200 E. California Blvd., Pasadena, CA 91125, USA}
\newcommand{\affRSAAO}{South African Astronomical Observatory, PO Box 9, Observatory 7935, South Africa}
\newcommand{\affRHeid}{Zentrum f{\"u}r Astronomie der Universit{\"a}t Heidelberg, Astronomisches Rechen-Institut, M{\"o}nchhofstr. 12-14, 69120 Heidelberg, Germany}
\newcommand{\affREdinburgh}{ Institute for Astronomy, University of Edinburgh, Royal Observatory,Edinburgh EH9 3HJ, UK}
\newcommand{\affRParis}{Sorbonne Universit\'e, CNRS, UMR 7095, Institut d'Astrophysique de Paris, 98 bis bd Arago, F-75014 Paris, France }

 \newcommand{\affSalerno}{Università degli Studi di Salerno, Dipartimento di Fisica “E.R. Caianiello”, via Ponte Don Melillo, 84085 Fisciano (SA), Italy}

\newcommand{\affMNU}{Institute for Space-Earth Environmental Research, Nagoya University, Nagoya 464-8601, Japan}
\newcommand{\affMGoddard}{Code 667, NASA Goddard Space Flight Center, Greenbelt, MD 20771, USA}

\newcommand{\affMaryland}{Department of Astronomy, University of Maryland, College Park, MD 20742, USA}
\newcommand{\affMUA}{Department of Physics, University of Auckland, Private Bag 92019, Auckland, New Zealand}

\newcommand{\affMOSA}{Department of Earth and Space Science, Graduate School of Science, Osaka University,Toyonaka, Osaka 560-0043, Japan}

\newcommand{\affMJAXA}{Institute of Space and Astronautical Science, Japan Aerospace Exploration Agency, Kanagawa 252-5210, Japan}
\newcommand{\affMCNY}{University of Canterbury Mt.\ John Observatory, P.O. Box 56, Lake Tekapo 8770, New Zealand}

\newcommand{\affCopenhagen}{Centre for ExoLife Sciences, Niels Bohr Institute, University of Copenhagen, Jagtvej 155, 2200 Copenhagen, Denmark}
\newcommand{\affMassey}{Institute of Natural and Mathematical Sciences, Massey University, Auckland 0745, New Zealand}
\newcommand{\affRoma}{Dipartimento di Fisica, Universit\`{a} di Roma Tor Vergata, Via della Ricerca Scientifica 1, 00133 Roma, Italy Italy}
\newcommand{\affMaxPHeidelberg}{Max Planck Institute for Astronomy, K\"{o}nigstuhl 17, D-69117, Heidelberg, Germany}
\newcommand{\affTurin}{INAF -- Turin Astrophysical Observatory, Via Osservatorio 20, I-10025, Pino Torinese, Italy}
\newcommand{\affCITEVA}{Centro de Astronom{\'{\i}}a (CITEVA), Universidad de Antofagasta, Avda.\ U.\ de Antofagasta 02800, Antofagasta, Chile}
\newcommand{\affCoimbra}{Instituto de Astrofísica e Ciências do Espaço, Universidade de Coimbra, 3040-004 Coimbra, Portugal}
\newcommand{\affKeynes}{Centre for Electronic Imaging, Department of Physical Sciences, The Open University, Milton Keynes, MK7 6AA, United Kingdom}
\newcommand{\affKeele}{Astrophysics Group, Keele University, Staffordshire, ST5 5BG, United Kingdom}
\newcommand{\affHamburg}{Universit{\"a}t Hamburg, Faculty of Mathematics, Informatics and Natural Sciences, Department of Earth Sciences, Meteorological Institute, Bundesstra\ss{}e 55, 20146 Hamburg, Germany}
\newcommand{\affPontificia}{Centro de Astroingenier\'ia, Facultad de F\'isica, Pontificia Universidad Cat\'olica de Chile, Av. Vicu\~na Mackenna 4860, Macul 7820436, Santiago, Chile}

\newcommand{\affSoutDenmark}{University of Southern Denmark, Department of Physics, Chemistry and Pharmacy, SDU-Galaxy, Campusvej 55, 5230 Odense M, Denmark}
\newcommand{\affManchester}{Jodrell Bank, School of Physics and Astronomy, University of Manchester, Oxford Road, Manchester M13 9PL, United Kingdom}
\newcommand{\affConcepcion}{Universidad Catolica de la Santisima, Concepcion, Chile}
\newcommand{\affTehran}{Department of Physics, Sharif University of Technology, PO Box 11155-9161 Tehran, Iran}
\newcommand{\affNapoli}{Istituto NJodrell Bank, School of Physics and Astronomy, University of Manchester, Oxford Road, Manchester M13 9PL, United Kingdomazionale di Fisica Nucleare, Sezione di Napoli, Napoli, Italy}
\newcommand{\affUTokyoEarth}{Department of Earth and Planetary Science, Graduate School of Science, The University of Tokyo, 7-3-1 Hongo, Bunkyo-ku, Tokyo 113-0033, Japan}
\newcommand{\affICanarias}{Instituto de Astrof\'{\i}sica de Canarias, V\'{\i}a Lacte\'{a} s/n, E-38205 La Laguna, Tenerife, Spain}
\newcommand{\affUTokyoAst}{Institute of Astronomy, Graduate School of Science, The University of Tokyo, 2-21-1 Osawa, Mitaka, Tokyo 181-0015, Japan}

\newcommand{\affOAK}{Oak Ridge Associated Universities, Oak Ridge, TN 37830, USA}
\newcommand{\affCatolicaAst}{Instituto de Astrof\'isica, Facultad de F\'isica, Pontificia Universidad Cat\'olica de Chile, Av. Vicu\~na Mackenna 4860, 7820436 Macul, Santiago, Chile}
\begin{document} 

\title{OGLE-2015-BLG-1609Lb: Sub-jovian planet orbiting a low-mass stellar or brown dwarf host}

\author{M.~J.~Mr{\'o}z\inst{\ref{OAUW}} \and R.~Poleski\inst{\ref{OAUW}} \and A.~Udalski\inst{\ref{OAUW}} \and 
T.~Sumi\inst{\ref{MOSA}}  \and
Y.~Tsapras \inst{\ref{RHeid}} \and
M.~Hundertmark \inst{\ref{RHeid}}
\\ (Leading Authors) \\ \vspace{0.2cm}
P.~Pietrukowicz \inst{\ref{OAUW}} \and
M.~K.~Szyma\'nski \inst{\ref{OAUW}} \and
J.~Skowron\inst{\ref{OAUW}} \and
P.~Mróz\inst{\ref{OAUW}} \and
M.~Gromadzki\inst{\ref{OAUW}} \and
P.~Iwanek\inst{\ref{OAUW}} \and
S.~Koz\l{}owski\inst{\ref{OAUW}}\and
M.~Ratajczak\inst{\ref{OAUW}} \and
K.~A.~Rybicki\inst{\ref{WEIZMANN},\ref{OAUW}} \and
D.~M.~Skowron \inst{\ref{OAUW}} \and
I.~Soszy\'nski \inst{\ref{OAUW}} \and
K.~Ulaczyk \inst{\ref{WARWICK}} \and
M.~Wrona \inst{\ref{Villanova}, \ref{OAUW}}
\\ (OGLE Collaboration) \\ \vspace{0.2cm}
F.~Abe  \inst{\ref{MNU}} \and 
K.~Bando \inst{\ref{MOSA}} \and 
D.P.~Bennett\inst{\ref{MGoddard}, \ref{Maryland} } \and 
A.~Bhattacharya\inst{\ref{MGoddard},  \ref{Maryland} } \and 
I.~A.~Bond \inst{\ref{Massey}}\and 
A.~Fukui \inst{\ref{UTokyoEarth}, \ref{ICanarias}} \and 
R.~Hamada \inst{\ref{MOSA}}\and 
S.~Hamada \inst{\ref{MOSA}}\and 
N.~Hamasaki  \inst{\ref{MOSA}}\and 
Y.~Hirao \inst{\ref{UTokyoAst}}\and 
S.~Ishitani~Silva\inst{\ref{MGoddard}, \ref{OAK}} \and 
Y.~Itow  \inst{\ref{MNU}} \and 
N.~Koshimoto \inst{\ref{MOSA}} \and  
Y.~Matsubara \inst{\ref{MNU}} \and 
S.~Miyazaki \inst{\ref{MJAXA}} \and 
Y.~Muraki\inst{\ref{MNU}} \and
T.~Nagai \inst{\ref{MOSA}} \and
K.~Nunota \inst{\ref{MOSA}} \and
G.~Olmschenk\inst{\ref{MGoddard}} \and 
C.~Ranc \inst{\ref{RParis}}\and 
N.~J.~Rattenbury  \inst{\ref{MUA}} \and 
Y.~Satoh \inst{\ref{MOSA}} \and 
D.~Suzuki \inst{\ref{MOSA}} \and 
S.~K.~Terry\inst{\ref{MGoddard}, \ref{Maryland} } \and 
P.~J.~Tristram\inst{\ref{MCNY}} \and 
A.~Vandorou\inst{\ref{MGoddard},  \ref{Maryland}}\and 
H.~Yama\inst{\ref{MOSA}} 
\\ (MOA Collaboration) \\ \vspace{0.2cm} 
R.~A.~Street\inst{\ref{RLCO}} \and 
E.~Bachelet \inst{\ref{RIPAC}} \and
M.~Dominik \inst{\ref{RSUPA}} \and 
A.~Cassan\inst{\ref{RParis}}\and 
R.~Figuera~Jaimes\inst{\ref{RMill}, \ref{CatolicaAst}}\and  
K.~Horne\inst{\ref{RSUPA}} \and 
R.~Schmidt \inst{\ref{RHeid}} \and
C.~Snodgrass \inst{\ref{REdinburgh}}\and
J.~Wambsganss\inst{\ref{RHeid}}\and
I.~A.~Steele\inst{\ref{RLiver}}\and
J.~Menzies\inst{\ref{RSAAO}}
\\ (RoboNet Collaboration) \\ \vspace{0.2cm} 
U.~G.~J{\o}rgensen \inst{\ref{Copenhagen}} \and 
P.~Longa-Pe{\~n}a \inst{\ref{CITEVA}} \and
N.~Peixinho \inst{\ref{Coimbra}} \and 
J.~Skottfelt \inst{\ref{Keynes}} \and 
J.~Southworth \inst{\ref{Keele}} \and 
M.~I.~Andersen\inst{\ref{RHeid}} \and 
V.~Bozza \inst{\ref{Salerno},  \ref{Napoli}}\and
M.~J.~Burgdorf\inst{\ref{Hamburg}} \and 
G.~D'Ago\inst{\ref{Pontificia}} \and 
T.~C.~Hinse\inst{\ref{SoutDenmark}} \and 
E.~Kerins\inst{\ref{Manchester}} \and 
H.~Korhonen\inst{\ref{RHeid}} \and 
M.~K\"{u}ffmeier\inst{\ref{RHeid}} \and 
L.~Mancini\inst{\ref{Roma}, \ref{MaxPHeidelberg}, \ref{Turin}} \and 
M.~Rabus\inst{\ref{Concepcion}} \and 
S. Rahvar\inst{\ref{Tehran}} 
\\(MiNDSTEp Collaboration)  \\ \vspace{0.2cm} }

\institute{ \affOAUW\label{OAUW} \and
\affMOSA \label{MOSA} \and 
\affRHeid \label{RHeid} \and
\affWEIZMANN \label{WEIZMANN} \and 
\affWARWICK \label{WARWICK} \and
\affVillanova \label{Villanova}  \and
\affMNU \label{MNU} \and 
\affMGoddard\label{MGoddard} \and
\affMaryland \label{Maryland} \and
\affMassey \label{Massey} \and
\affUTokyoEarth\label{UTokyoEarth} \and 
\affICanarias\label{ICanarias}  \and 
\affUTokyoAst\label{UTokyoAst} \and 
\affOAK\label{OAK} \and 
\affMJAXA\label{MJAXA} \and 
\affRParis \label{RParis} \and
\affMUA \label{MUA} \and 
\affMCNY \label{MCNY}\and 
\affRLCO \label{RLCO} \and 
\affRIPAC \label{RIPAC} \and 
\affSalerno \label{Salerno} \and 
\affNapoli \label{Napoli} \and 
\affRSUPA \label{RSUPA}\and  
\affRMill \label{RMill}\and 
\affCatolicaAst\label{CatolicaAst}\and  
\affREdinburgh \label{REdinburgh} \and 
\affRLiver \label{RLiver} \and 
\affRSAAO \label{RSAAO} \and  
\affCopenhagen \label{Copenhagen} \and 
\affCITEVA \label{CITEVA} \and 
\affCoimbra\label{Coimbra} \and 
\affKeynes\label{Keynes} \and 
\affKeele \label{Keele} \and 
\affHamburg \label{Hamburg} \and 
\affPontificia \label{Pontificia} \and  
\affSoutDenmark\label{SoutDenmark} \and 
\affManchester\label{Manchester} \and  
\affRoma\label{Roma} \and  
\affMaxPHeidelberg \label{MaxPHeidelberg} \and
\affTurin \label{Turin} \and
\affConcepcion\label{Concepcion} \and  
\affTehran\label{Tehran} }
\date{Received December XX, 20XX}

  \abstract{We present a comprehensive analysis of a planetary microlensing event OGLE-2015-BLG-1609. The planetary anomaly was detected by two survey telescopes, OGLE and MOA. Each of these surveys  collected enough data over the planetary anomaly to allow for an unambiguous planet detection. Such survey detections of planetary anomalies are needed to build a robust sample of planets that could  improve studies on the microlensing planetary occurrence rate by reducing biases and statistical  uncertainties. In this work, we examined different methods for modeling microlensing events using individual datasets, particularly we incorporated a Galactic model prior to better constrain poorly defined microlensing parallax. Ultimately, we fitted a comprehensive model to all available data, identifying three potential typologies, with two showing comparably high Bayesian evidence.
  Our analysis indicates that the host of the planet is a brown dwarf with a probability of 34\%, or a low-mass stellar object (M-dwarf) with the probability of 66\%. 
}

   \keywords{gravitational lensing: micro -- planets and satellites: detection}

   \maketitle

\section{Introduction}
More than three decades have passed since \cite{1991ApJ...374L..37M} first suggested that extrasolar planets could be detected in  microlensing events. During this time, over 230 planets\footnote{\url{https://exoplanetarchive.ipac.caltech.edu}} have been discovered using this method.  This number may seem small compared to discoveries made by other major planet-detection techniques, but the strength of microlensing lies in its unique sensitivity to low-mass and distant planets. Instead of analyzing the light of the planet's host star, microlensing uses the light of an unrelated background star to probe the foreground planetary system. This technique allows for the access to ranges of parameters, such as orbital sizes and host distances from the Galactic center, which are currently not achievable by any other method \citep{2018Geosc...8..365T}.\par
Another advantage of microlensing is its powerful capability of demographic studies\footnote{\url{https://exoplanets.nasa.gov/internal_resources/2749/ExEP_Science_Gap_List_2023_Final.pdf}}. Typically, planetary occurrence rates in microlensing are derived as a function of mass ratios ($q$) and projected separations in the Einstein ring radius units ($s$), calculated using the set of events and the detection efficiency of the survey that detected them \citep{Mroz2020}. However, these analyses are affected by uncertainties and biases that require significant effort to disentangle (e.g., \citealt{2010ApJ...720.1073G}, \citealt{2018AcA....68....1U}). These factors are inherent in the sample of analyzed microlensing events, starting with observational data collected by numerous surveys using various instruments and under differing conditions. Furthermore, variations in data reduction and selection processes, along with subjective human factors such as  publication bias \citep{2020AJ....159...98Y}, contribute to these uncertainties. Collecting a sample of events detected from homogeneous data gathered by a single instrument greatly minimizes these issues.\par
In this work, we present the discovery of the planetary microlensing event OGLE-2015-BLG-1609. The planetary signal is detectable by the survey data alone, which allows for including this planet in the statistical analysis of microlensing planets based solely on the survey data. The analysis turned out to be more complex than typical, due to low-level systematic trends in the photometry. We found that there are three different topologies that could explain the light curve of the event and two of them have very similar Bayesian evidence, hence, we were not able to distinguish between them. We estimated the mass of the planet host, which overlaps with masses of both stars and brown dwarfs. \par
One of the most detailed studies on the planetary occurrence rate in microlensing events was done by the Microlensing Observations in Astrophysics (MOA) collaboration, using their 2007–2012 sample \citep{2016ApJ...833..145S}. This sample comprised 1,474~events, including 23 planetary events. \citet{2016ApJ...833..145S} found that the planetary occurrence rate follows a broken power law of~$q$. 
However, due to the lack of events in the sample with $q < 10^{-4.5}$, which is close to the break in the power law, fitting the value $q_{\rm{br}}$ led to high uncertainty in all the parameters. Therefore, the authors fixed this value at $q_{\rm{br}} = 1.7 \times 10^{-4}$ and concluded that beyond the snow line, most planets should have Neptune-like masses.  
Later,  known planets with $q<10^{-4}$ were analyzed by
\cite{2018AcA....68....1U}. Combining planets detected using both survey and follow-op data forced authors to apply an alternative inference method ("$V/V_{\rm{max}}$" method; \citealt{1968ApJ...151..393S}). \cite{2018AcA....68....1U} confirmed the break in the mass ratio power law but at the higher masses, $q_{\rm{br}} = 2 \times 10^{-4}$. 
The break was further analyzed by \cite{2019AJ....157...72J}, using a sample of 15 planets with $q < 3 \times 10^{-4}$ and assuming that planet-detection sensitivity as a function of $q$ can be approximated by a simple power law. Their results suggest that the break is at a much smaller mass ratio ($q_{\rm{br}} \approx 0.55 \times 10^{-4}$) and is much steeper compared to the findings by \cite{2016ApJ...833..145S}. \par
Currently, significant efforts are undertaken, using the Korea Microlensing Telescope Network's (KMTNet) semi-automated algorithm, \texttt{Anomalyfinder} \citep{2021AJ....162..163Z}, to build a large, uniformly selected sample of planetary events. So far, the algorithm has identified about 100 planets in KMTNet's observations from the years 2016-2019  (\citealt{2021AJ....162..163Z};
\citealt{2022AJ....163...43H};
\citealt{2022MNRAS.510.1778W};
\citealt{2022A&A...664A..13G};
\citealt{2022MNRAS.515..928Z};
\citealt{2022AJ....164..262J};
\citealt{2023AJ....165..103Z};
\citealt{2023AJ....165..226J};
\citealt{2023AJ....166..104S};
\citealt{2024AJ....167...88R};
\citealt{2024AJ....167..269S}), including a few with $q<0.5 \times 10^{-4}$. Future statistical analysis of this sample will significantly improve constraints on planetary occurrence rates, in particular in the low-mass regime.
Projected separation and mass ratio are not the only parameters that can be derived from microlensing events.
By measuring the so-called secondary effects in the light curve, it is possible to obtain absolute masses, the distance to the system, and the projected separation in physical units. Typically, for this purpose, a pair of finite-source and microlensing parallax effects is used. Despite the fact that both these effects are measured only in  $ \sim 20 \%$ of planetary events, some initiatives were taken to measure how the distribution of the planets depends on the host mass and Galactocentric distance ($M_{\rm{L}},~R_{\rm{L}}$).
The hypothesis of equal abundance of planets in the bulge and the disc was tested by \cite{2016ApJ...830..150P}, using a sample of 21 systems.
However, the small sample of the parallax measurements, hindered by systematic uncertainties, prevented them from reaching a definitive conclusion. 
Additionally, some nearby planets (< 2 kpc) in the sample have now been shown, through analysis of high-angular-resolution images, to be at larger distances than originally reported (e.g., MOA-2007-BLG-192 -- \citealt{2024AJ....168...72T})).
Using a slightly larger sample (28 planets), \cite{2021ApJ...918L...8K} calculated the planetary occurrence rate as a function of both host mass and Galactocentric distance, $P_{\rm{host}}~\propto M^{{m}}_{\rm{L}}  R^{{r}}_{\rm{L}}$, by comparing the observed distribution of lens-source relative proper motion ($\mu_{\rm{rel}}$) to Galactic model predictions for a given Einstein radius crossing time ($t_{\rm{E}}$).
Their results suggest that the probability of hosting a planet increases with the Galactocentric distance, but this dependence is not statistically significant, with $r=0.2 \pm0.4$.
The uncertainties in the exponents of  $R_{\rm{L}}$ and $M_{\rm{L}}$  were reduced by \cite{2024ApJ...967...77N} by utilizing model distributions of both $\mu_{\rm{rel}}$ and $t_{\rm{E}}$ and dividing systems into two subsamples with mass ratios below and above $q = 10^{-3}$. 
Their analyses indicate that beyond the snow line, massive planets are more likely to be around more massive stars, while the frequency of low-mass planets does not depend on the mass of the host.\par
The planetary event analyzed here will be included in the statistical studies of
MOA or OGLE survey planets. Such studies can shed more light on the detailed
aspects of the planetary occurrence rate.\par
This paper is organized as follows: In Section 2, we describe the analysis of the observational data. In Section 3, we detail the model fitting to the OGLE observations alone. In Section 4, we present the final results of modeling using all available data sets and the process of estimating the physical parameters of the system. Finally, we summarize our results in Section 5.
\section{Observational data}
\subsection{Collection}
The microlensing event OGLE-2015-BLG-1609 was first alerted by the Optical Gravitational Lensing Experiment (OGLE) on 2015 June 16, Heliocentric Julian Date (HJD) $\sim 2457219$. The detection was made by the Early Warning System (EWS \footnote{\url{https://ogle.astrouw.edu.pl/ogle4/ews/ews.html}}; \citealt{2003AcA....53..291U}) in the data from  the 1.3-m Warsaw Telescope, equipped with the $1.4~\rm{deg}^2$ field of view mosaic CCD camera at the Las Campanas Observatory in Chile \citep{2015AcA....65....1U}. The  event was located at $(\rm{RA},\rm{Dec})_{2000}$ = (18:03:17.71,~-26:54:25.6) in equatorial coordinates, or $(l,b) = (3.^\circ72,-2.^\circ34)$ in Galactic coordinates, which is inside the BLG511 field of the OGLE-IV survey. This field was one of the OGLE high cadence Galactic Bulge fields, observed 4 times per night on average. OGLE images were mostly taken in the $I-$band, except for the observations in the $V$ band, taken once per several days. The reduction of the OGLE data was done using a variant of difference image analysis (DIA;\citealt{1996ApJ...473L..87C, 1998ApJ...503..325A}) optimized by \cite{2000AcA....50..421W} and \cite{2008AcA....58...69U}.\par
The event was also observed by MOA's 1.8-m telescope at Mt. John University Observatory in New Zealand. It was independently alerted on 2015 June 30, HJD $\sim 2457233$, as MOA-2015-BLG-412 on the MOA alerts webpage\footnote{\url{https://www.massey.ac.nz/~iabond/moa/alerts/}}. MOA observations were primarily done in the custom wide MOA-Red filter, approximately equal to the sum of $R-$ and $I-$ band filters. MOA data reductions were performed using a variation of DIA presented by \cite{2001MNRAS.327..868B}.
Additionally, MOA observed in the $V$ band, but these few observations were excluded in this analysis due to large error bars.\par
\begin{figure*}[htb!]
    \centering
    \includegraphics[width=1\linewidth]{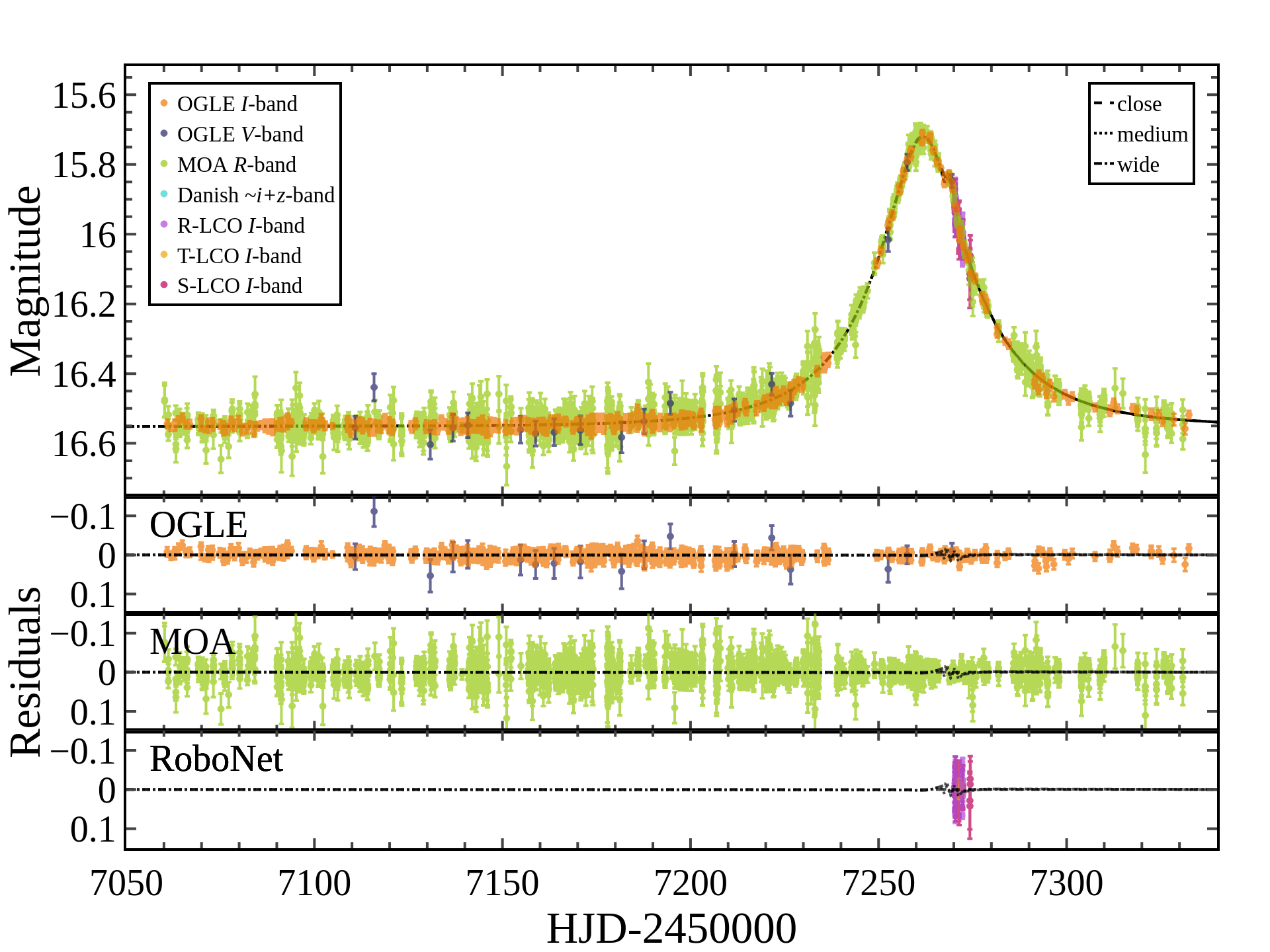}
    \caption{Light curves of the microlensing event OGLE-2015-BLG-1609 with three microlensing model topologies for positive values of the impact parameter $u_{\rm{0}}$. The difference between the topologies is visible only during the planetary anomaly (HJD $\sim 2457270$; see Fig. \ref{fig:modle_zoom}).  The photometry comes from the OGLE, MOA, RoboNet, and MiNDSTEp projects.}
    \label{fig:model}
\end{figure*}
\begin{figure}[htb!]
    \centering
    \includegraphics[width=1\linewidth]{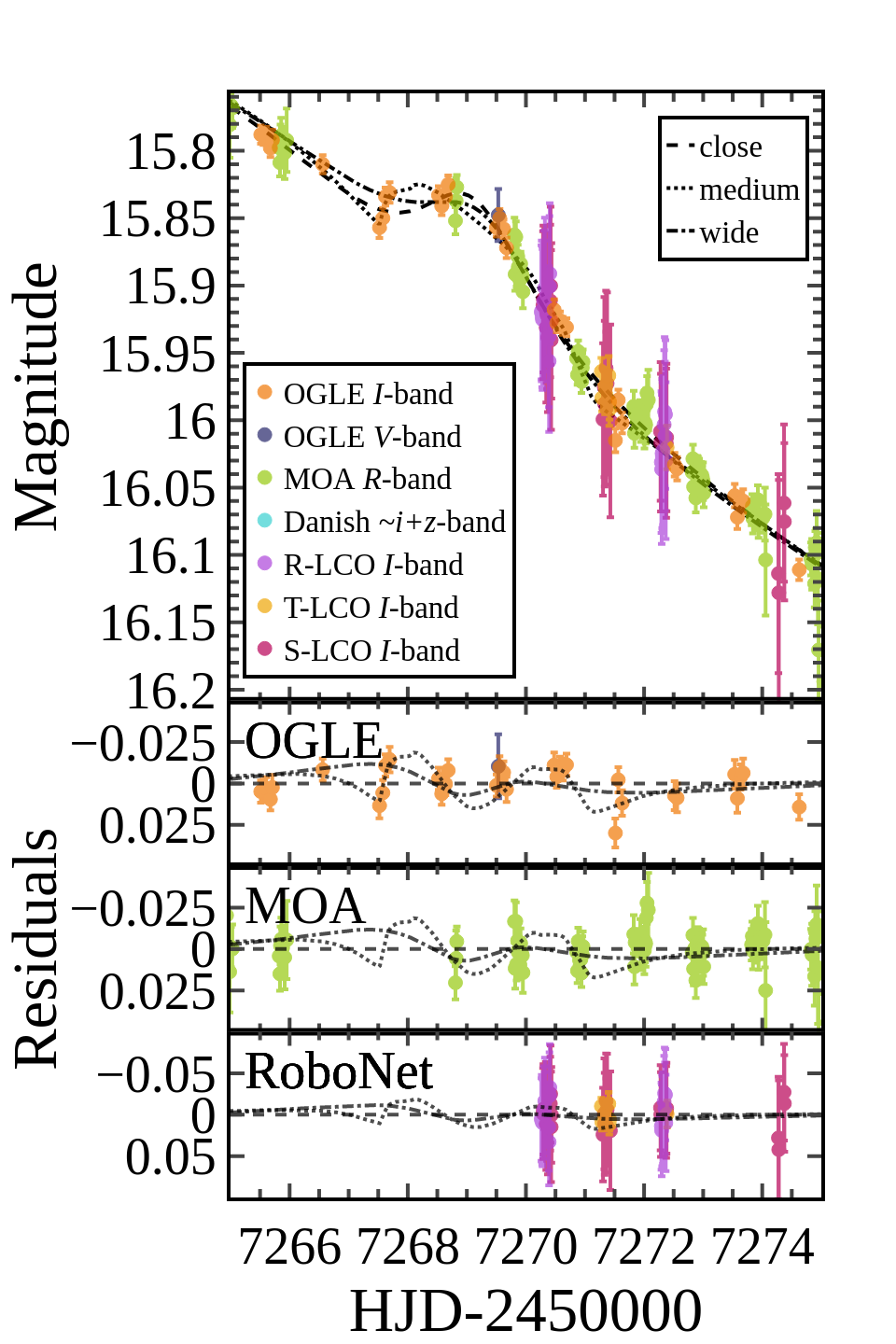}
    \caption{Light curve of the planetary anomaly in the microlensing event OGLE-2015-BLG-1609 with three microlensing model topologies for positive values of the impact parameter $u_{\rm{0}}$. The photometry comes from the OGLE, MOA, RoboNet, and MiNDSTEp projects.  }
    \label{fig:modle_zoom}
\end{figure}

On 2015 September 3, HJD $\sim  2457269$, the light curve visibly deviated from the standard  \cite{1986ApJ...304....1P} model, indicating the presence of a planetary anomaly. This triggered follow-up observations on three 1-m Las Cumbres Observatory (LCO) telescopes located at the South African Astronomical Observatory, operated by the RoboNet collaboration \citep{2009AN....330....4T}, and on the 1.54 m Danish Telescope at La Silla Observatory, Chile, operated by the Microlensing Network for the Detection of Small Terrestrial Exoplanets\footnote{\url{http://www.mindstep-science.org}}  collaboration (MiNDSTEp; \citealt{2010AN....331..671D}). 
RoboNet data reductions were performed using a modified DIA algorithm (\texttt{DANDIA}; \citealt{2008MNRAS.386L..77B}). 
Observations on the Danish Telescope were collected using a high frame-rate electron-multiplying CCD (EMCCD; \citealt{2015A&A...574A..54S}) and were reduced using a variation of the \texttt{DANDIA} algorithm optimized for this instrument.
The light curve of the event is plotted in Figure~\ref{fig:model} and zoomed in on the planetary anomaly in Figure~\ref{fig:modle_zoom}.
Some data were collected using other telescopes but due to poor coverage of the event, large error bars, or the lack of  reliable reductions, we excluded them from this analysis. 
\subsection{Primary analysis}
In the initial analysis of  OGLE-2015-BLG-1609 based solely on the OGLE survey data, we faced several challenges.
In models that include the microlensing parallax effect, we obtained well-constrained but astrophysically unlikely values for the northern component of the parallax vector ($\pi_{\mathrm{E},N}$), which should be close to zero according to the Galactic models (e.g.,  \citealt{2020ApJ...889...31L}, \citealt{2021ApJ...917...78K}). Additionally, at the end of the event and before the annual break in the visibility of the Galactic bulge, a trend at HJD 2457300 -- 2457335  was noticed in the residuals from the best fit. We tried to solve these issues by including additional effects in microlensing models. First considered was the binary source effect (2L2S). In this model, the brightening at the end of the event was caused by the presence of a second star in the source system. Inclusion of this effect improved $\chi^2$ of the fit, but did not change the value of $\pi_{\mathrm{E},N}$. Since only half of the second source brightening was covered by data points, we concluded that this trend is more likely an end-of-the-season effect than an astrophysical one, and we rejected the binary source explanation. \par
We also considered additional changes in the geometry of the event caused by either the orbital motion of the lenses or the orbital motion of the source (xallarap effect). In the case of lens orbital motion, we did not achieve reasonable constraints on the effect's parameters, nor did we improve the $\chi^2$ statistics of the fits. This can be explained by the lack of well-timed light-curve features, such as a caustic crossing \citep{2001ApJ...563L.111A}. Similarly, in models with the xallarap effect, there was no improvement in the $\chi^2$ statistics. Additionally, the resulting values of the orbital period of the source aligned with spurious periods of low-level amplitude recurring in the OGLE photometric data, as found in previous studies \citep{2023AcA....73..127M}. 
These periods are associated with the structure of the data itself. For this reason, we do not consider the results of xallarap models to be trustworthy.
Including lens orbital motion or xallarap effects in the models did not change the values of  $\pi_{\mathrm{E},N}$.
Overestimation of the parallax vector may be caused by other unassociated photometric variability.
To address the potential trends in our data we incorporated Gaussian processes into modeling of the event. 
This approach has been successful in modeling the variability of the microlensing source, demonstrated in the analysis of the event OGLE-2017-BLG-1186 \citep{2019MNRAS.488.3308L}.
Unfortunately, it led to an increase of the $\pi_{\mathrm{E},N}$ value with even smaller uncertainties.
Simultaneous modeling of the variability of the source star and the parallax effect requires a mindful choice of the time range of the baseline used. 
A long baseline is needed to reliably constrain the long-term stellar variability. However, extending the baseline much beyond the duration of the microlensing event can adversely affect parallax vector measurements by introducing uncorrelated trends. \par
Finally, to ensure the absence of long-term observational trends in the light curve of the event, we  analyzed observations of the nearby stars in the OGLE database. We selected 16 constant stars in the OGLE database around $30\arcsec$ from the source star, with similar brightness and color. We arranged their observations in week-long bins, then calculated the average between all the chosen stars. The resulting trend did not exceed the average magnitude uncertainty of the analyzed data, thus could not affect the microlensing modeling. \par
After exhausting the approaches detailed above, we decided to include in our analysis the data collected by the MOA group. Models fitted to the MOA data alone and to the combined MOA and OGLE datasets provided more typical results for the $\pi_{\mathrm{E},N}$ value, which was consistent with zero within the $2 \sigma$ limit. This led us to make a step back and take a closer look at the reductions of the OGLE images. In the DIA reduction method, photometric measurements are obtained by subtracting a reference image from the acquired images. The reference images are created by averaging a selected dozen or so images taken under the best seeing conditions. In the case of the initial OGLE reductions, reference images were taken at the beginning of OGLE-IV survey, in 2010. Three years after the event, in 2018, the OGLE team prepared new reference images. Since they were closer in time to the analyzed event, we used them to re-extract the photometry. We fitted the model with parallax effect to this new OGLE dataset. This time we obtained a distribution of $\pi_{\mathrm{E},N}$ that was consistent with zero within the $2 \sigma$ limit, therefore, in a good agreement with the Galactic model. The photometry extracted using DIA can be affected at a low level by the proper motion of the source star. Hence, changes in the position of the source star can influence microlensing parallax measurements. We present in Table~\ref{tab:PM} proper motion measurements from the Gaia DR3 \citep{2023A&A...674A...1G} and the internal OGLE time-series astrometry, which has been matched to the Gaia DR3 reference frame. Note that the uncertainties in the OGLE measurements are purely statistical and likely underestimated. 
The value of the renormalized unit weight error (RUWE > 1.5; Table~\ref{tab:PM}) for the Gaia DR3 asymmetric solution indicates that the parallax measurements from Gaia DR3 cannot be attributed to either the lens or the source in the event \citep{2023A&A...674A..23W}.\par
In all cases, the binary-lens models fitted to individual datasets from both the "old" and "new" OGLE reductions, as well as  MOA observations, have significantly lower $\chi^2$ statistics compared to the single-lens models fitted to the same datasets. This enables the inclusion of this event in the planetary occurrence rate studies of both surveys. The comparison is shown in the Table~\ref{tab:chi2}.
In the subsequent investigation, we used "new" OGLE reductions only.
\begin{table}
\caption{Proper motion of the source in the OGLE-2015-BLG-1609 event.}
\label{tab:PM}
\centering 
\begin{tabular}{c|c|c}
\hline\hline         
Database & $\mu_\alpha$ [mas] & $\mu_\delta$ [mas]\\
\hline        
Gaia DR3\tablefootmark{a} 
 & $-3.54 \pm   0.45$ & $   -5.53   \pm0.31$ \\
 OGLE   &$-2.66 \pm0.11$ & $ -6.24  \pm  0.05$\\
\hline 
\end{tabular}\tablefoot{
\tablefoottext{a}{RUWE=2.62}}
\end{table}
\begin{table*} 
\caption{Comparison of $\chi^2$ statistics for single-lens single-source (1L1S) and binary-lens single-source (2L1S) models, fitted separately to data from the "old" (based on the reference image from 2010) and "new" (based on the reference image from 2018) OGLE reductions, as well as MOA observations.} \label{tab:chi2}
\centering 
\begin{tabular}{c|c|c|c}
\hline\hline  
 &OGLE "old" reductions&OGLE "new" reductions&MOA \\
\hline 
$\chi^2/\rm{d.o.f}(1L1S)$ & $1170.38/842$& $1271.28/824$& $2653.70/2426$ \\
$\chi^2/\rm{d.o.f}(2L1S)$&$763.55/846$& $809.79/828$&  $2417.47/2430$ \\
\hline
$\Delta\chi^2$&$406.8$&$461.5$&$236.2$\\
\hline 
\end{tabular}
\end{table*}
\subsection{Data preparation and photometric uncertainties}
We performed  the color calibration of the OGLE $I-$ and $V-$band measurements,  so that magnitudes reported in this work are in the standard $I$ (Cousins) and $V$ (Johnson) pass-bands.
To clean datasets, we excluded all observations with error bars larger than five times the median of the error bars of nearby points. 
Additionally, using one of the early models of the event, we removed $3\sigma$ outliers from the fitted light curve. \par
Photometric pipelines sometimes struggle with estimating the impact of systematics in the data; uncertainties in the raw measurements are often underestimated and have broader tails than a Gaussian. 
In order to renormalize photometric error bars, we scaled them using the formula for the $i$-th data point \citep{2012ApJ...755..102Y}: 
\begin{equation}
    \sigma_{\mathrm{new}, i}=\sqrt{(k \sigma_i)^2 +e_{\rm{min}}^2},
        \label{equ:err}
\end{equation}
where $\sigma_i$ is an initial error bar, $k$ and $e_{\rm{min}}$ are renormalization coefficients.
Since we used new reference images for OGLE data reductions, the values of $k$ and $e_{\rm{min}}$ from the empirical model for this dataset reported by \cite{2016AcA....66....1S} could be inaccurate.
For the sake of coherent treatment of each dataset, we regarded $k$ and $e_{\rm{min}}$ for each dataset as additional parameters of the model and sampled with Markov Chain Monte Carlo (MCMC). We assumed that error bars should be Gaussian in flux space, and used the modified likelihood function in the form of: 
\begin{equation}
\begin{aligned}
  \ln{\mathcal{L}} & = -\frac{1}{2} \sum^N_{i=1} \left[ \left(\frac{f_i-f_{\mathrm{mod},i}}{\sigma_{\mathrm{new},i}} \right)^2 + \ln ( 2 \pi \sigma_{\mathrm{new},i}^2) \right] \\ & =   -\frac{1}{2} \left[ \chi^2 +  \sum^N_{i=1} \ln ( 2 \pi \sigma_{\mathrm{new},i}^2 )\right],
   \end{aligned}
\end{equation}
where $f_i-f_{\mathrm{mod},i}$  represents the difference between the measured and modeled flux values at the $i$-th observation.
All the modeling presented in this work was initially done using this approach. Then, the error bars scaling parameters were fixed to the values from the models with the lowest $\chi^2$, and the modeling was repeated.
\section{Single instrument data modeling -- OGLE}
As the planetary anomaly is well-sampled in the OGLE observations alone, initially we focused on just this dataset in our analysis.
The observed brightening in the OGLE-2015-BLG-1609 light curve, except for the anomaly, can be  described  well by the classical \cite{1986ApJ...304....1P} model. In this single-lens point-source model (1L1S), microlensing magnification is described by:
\begin{equation}
    A(u)=\frac{u^2+2}{u\sqrt{u^2+4}},
\end{equation}
$u(t)$ is projected separation between source and lens at the given time in units of the angular Einstein radius  $\theta_{\rm{E}}$  given by:
\begin{equation}
u(t)= \sqrt{u_0^2 + \tau(t)^2}      \quad\mathrm{and}\quad  \tau(t)=\frac{t-t_0}{t_{\rm{E}}} ,
\end{equation}
where $u_0$ is the impact parameter at the time of the closest approach between the source and the lens center of mass -- $t_0$. 
Taking into account that not all the measured light is magnified, the flux change as a function of time is given by:
\begin{equation}
    f(t)=f_{\rm{S}} A +f_{\rm{B}},
\end{equation}
where $f_{\rm{S}}$ and $f_{\rm{B}}$ are the fluxes of the source and blended objects, respectively. As the starting point of our analysis, we excluded data points collected during the anomaly (HJD 2457267 -- 2457273) and fitted the Paczyński curve to the remaining observations.  The achieved values of $t_{\rm{E}}$, $t_0$, and $u_0$ were used as starting values in further modeling. 
In the next step, we considered the binary-lens single-source (2L1S) model described in Section 3.1.  We extended this model by including the microlensing parallax effect (Section 3.2). Moreover, due to unreliable parameter constraints, we imposed an additional prior on 
 $\pi_\mathrm{E}$ from the Galactic model, as described in Section 3.3. \par
All the microlensing modeling in this work was done using the \texttt{MulensModel} Python-code \citep{2019A&C....26...35P}, with the Affine Invariant MCMC Ensemble sampler \citep{2010CAMCS...5...65G} implemented in the \texttt{emcee} code \citep{2013PASP..125..306F}. 
The number of chains and steps for each MCMC modeling run was adjusted to achieve the convergence. All runs with fixed error bars scaling parameters had a mean acceptance fraction between 0.287 and 0.458, the number of steps is at least 34 times longer than the mean autocorrelation time, and no thinning was applied (for definitions, see \citealt{2013PASP..125..306F}).
\subsection{Binary-lens model} 
The  planetary anomaly  in the light curve led us to consider a binary-lens model with finite-source effect (2L1S). This model requires extending the number  of parameters from three in Paczyński's model to seven. Those  additional parameters are typically:  $s$ -- the projected separation between the primary lens and the companion, $q$ -- the  ratio of the companion's mass to the primary lens mass, $\alpha$ -- the angle between the source trajectory and the binary axis, and the parameter of the finite source effect: $\rho$ -- the radius of the source in the Einstein radius units. In our modeling, we used this parameterization. Other alternative parameterizations were not suitable, as most of them are connected to features of the light curve during caustic crossings, which the analyzed event lacks. \par
\begin{figure}[htb!]
    \centering
    \includegraphics[width=1.\linewidth]{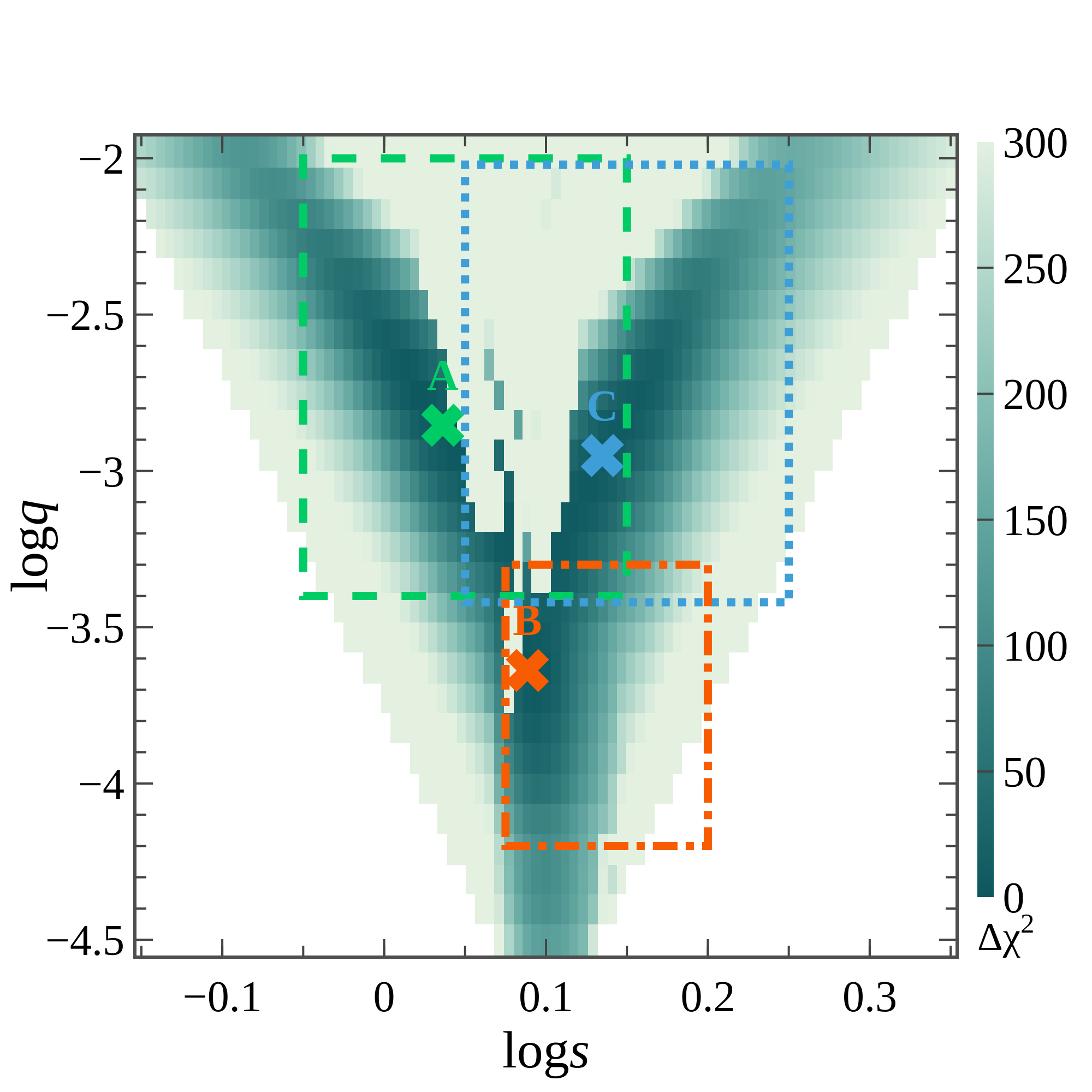}
    \caption{${\Delta\chi^2}$  map in $(\log{s}, \log{q})$ parameter space for OGLE-2016-BLG-1609 event. We named three visible topologies according to the projected separation: “close”--"A", dashed green  line; “medium” -- "B", dot-dashed orange line; and "wide" -- "C",  dotted blue line. Crosses mark the lowest $\chi^2$ models for each topology and boxes represent the $s$ and $q$ limits used in modeling.}
    \label{fig:heatmapchi2}
\end{figure}

To find all the degenerate solutions, we conducted a grid search in the $\log s-\log q$ parameters space. We used fixed values of $s$ evenly spaced between  $-0.15
<\log s<0.35$ and $q$ spaced between $-4.5<\log q<-2$, and optimized with MCMC method values of other parameters ($t_E$, $t_0$ $u_0$, $\rho$, $\alpha$). As the absence of a visible caustic crossing results in  poor constraints on $\alpha$, we split our modeling into two steps. First, we spread uniformly starting values of $\alpha$ and ran \texttt{EMCEE} with 300 walkers. After confirming that all models across $\log s-\log{q}$ grid gave consistent constraints on $\alpha$, we repeated modeling with a narrowed  starting range of $\alpha$.
We show the resulting $\chi^2$ map in Figure~\ref{fig:heatmapchi2}. \par
\begin{figure}[htb!]
    \centering
    \includegraphics[width=1\linewidth]{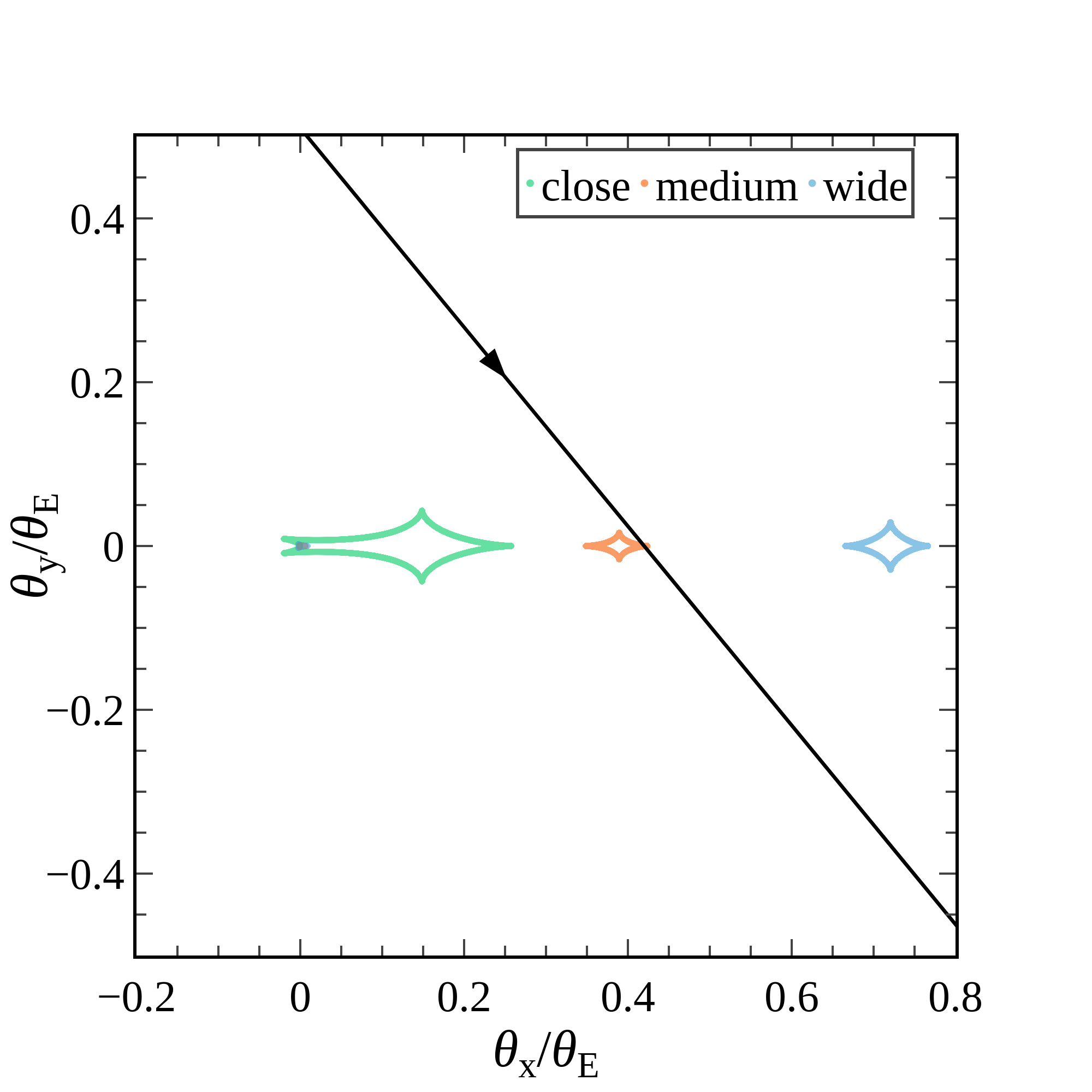}
    \caption{Source trajectory and caustics of OGLE-2015-BLG-1609 for different binary lens topologies: “close”-- green line, “medium” -- orange line, and "wide" -- blue line. The central caustics of the "medium" and "wide" models at the point $(\theta_{\rm{x}}/\theta_{\rm{E}},\theta_{\rm{y}}/\theta_{\rm{E}})=(0,0)$ are approximately point-like. For clarity, we plotted only the "middle" topology trajectory, as the other trajectories differed only slightly.}
    \label{fig:traj}
\end{figure}
We identified three possible topologies, depending on the point where the trajectory of the source crosses the binary axis. The “close” topology is in the case in which the trajectory passes behind the planetary caustic with respect to the central caustic; “wide”, in which the trajectory crosses between the central and the planetary caustic; and an intermediate case, “medium”, in which the trajectory passes through the planetary caustic, possible only in the case of a small mass ratio between the lenses (Figure~\ref{fig:traj}).
Finally, we fitted each of the three topologies separately, with all parameters freed.  For $s$ and $q$ the starting values were taken from the lowest $\chi^2$ models and these parameters were constrained to the ranges marked in Figure~\ref{fig:heatmapchi2}.
\subsection{2L1S model with the parallax effect} 
To obtain an estimate of the absolute values of the lens masses, additional effects in the light curve have to be considered. Typically, the most prominent effect is the microlensing parallax. It is a deviation from the straight trajectory of the source in the lens plane caused by the Earth's orbital motion. This effect can be used as a source of information on the relative movement of the source and the lens. We introduced this effect into our models by adding  North and East components of the microlensing parallax vector, $\pi_{\mathrm{E},N}$ and $\pi_{\mathrm{E},E}$, as parameters. \par
\begin{figure*}[htb!]
    \centering
    \includegraphics[width=1\linewidth]{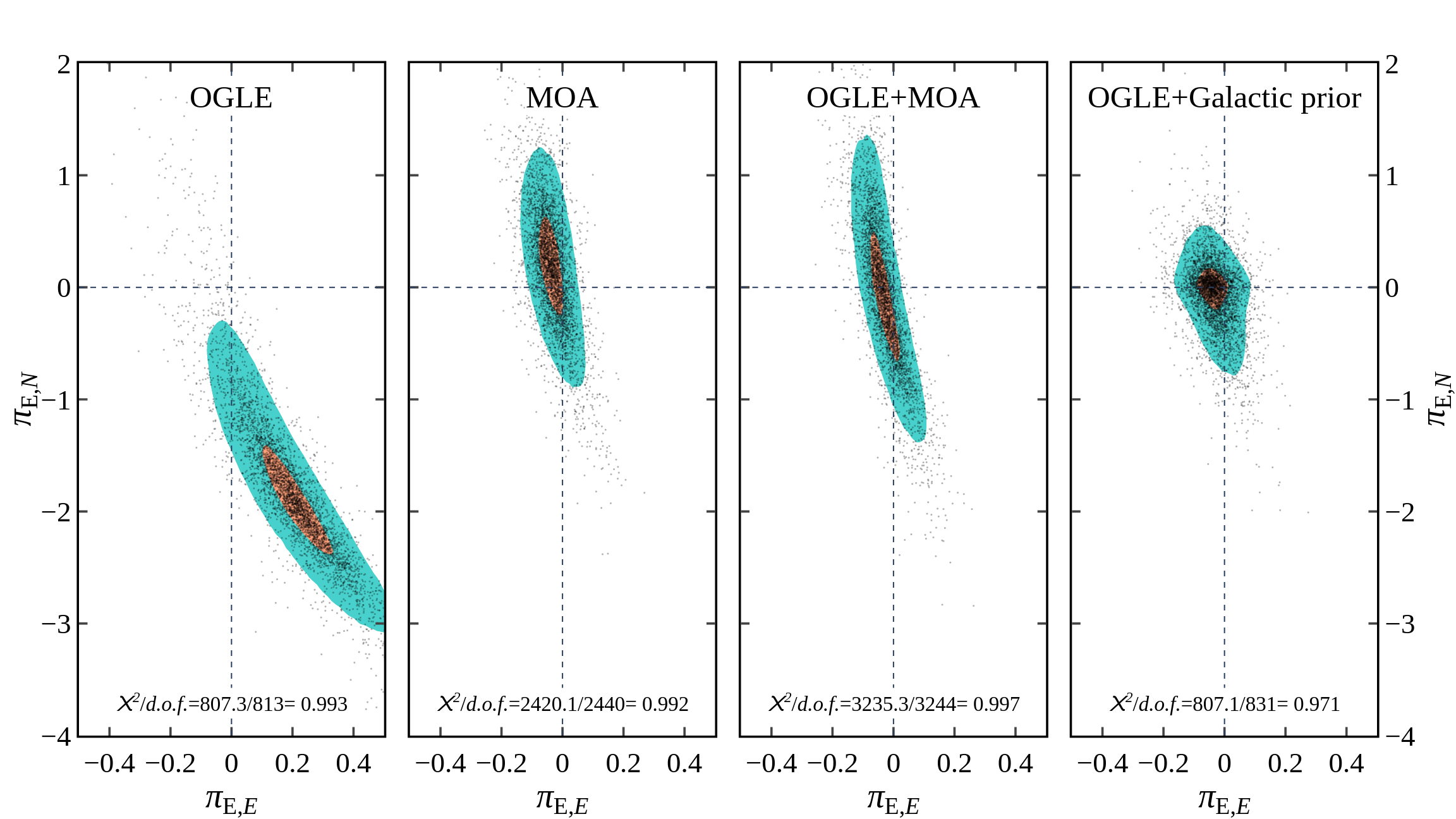}
    \caption{Comparison of the posterior distribution of the microlensing parallax vector $\pi_{\mathrm{E}}$ for "wide $u_{\rm{0}}+$" topology fitted to different combinations of datasets, from left: only OGLE data, only MOA data, combined OGLE and MOA datasets, and OGLE data alone with an incorporated prior on $\pi_{\mathrm{E}}$ based on a Galactic model. The orange region marks the $1\sigma$ contour of the two-dimensional distribution, and the turquoise region marks the $2\sigma$ contour.}
    \label{fig:PI_E}
\end{figure*}
\begin{figure}[htb!]
         \centering
         \includegraphics[width=1\linewidth]{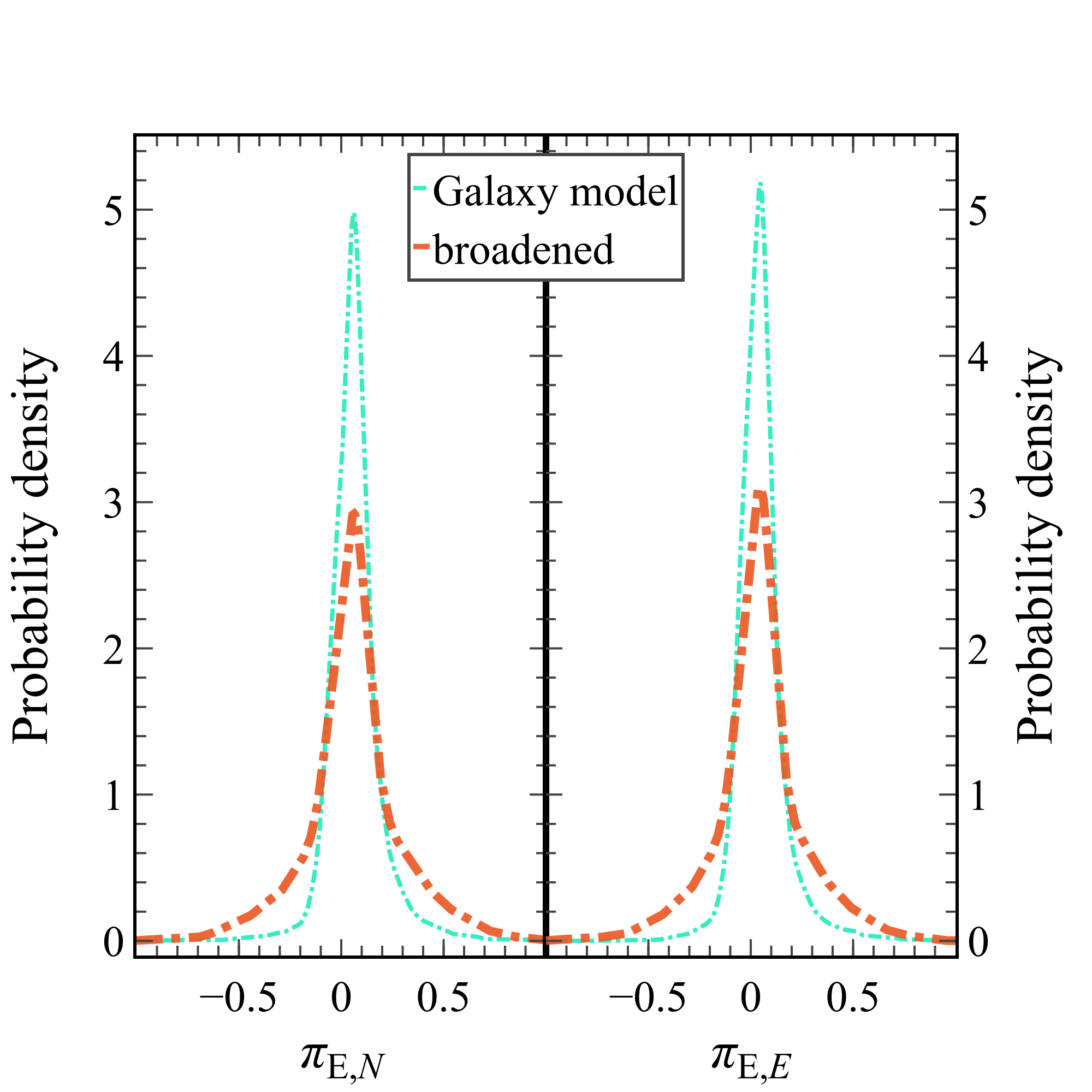}
         \caption{Probability density functions of the two components of the microlensing parallax vector ($\pi_{\mathrm{E},N}$ in the left panel and $\pi_{\mathrm{E},E}$ in the right panel), calculated from the Galactic model \citep{2021ApJ...917...78K}. The  dash-dotted turquoise line represents the original function from the simulated events, while the dash-dotted orange  function is broadened. The broadened functions are used as priors in the Bayesian inference.}
         \label{fig:pdf_PI_E}
     \hfill
\end{figure}

Because the parallax effect breaks the symmetry between positive $u_0$ and negative $u_0$ topologies, we additionally split our models between positive and negative $u_0$ values. 
The introduction of the parallax effect lowered the $\chi^2$  statistic of our models. However, the obtained values of the North component of the parallax vector were notably low ($\pi_{\mathrm{E},N} = -1.85^{+0.61}_{-0.45}$, for the "wide" $u_0+$ topologies; as shown in Figure~\ref{fig:PI_E}). These results contrast with the predictions from the Galactic models, which suggest $\vert \pi_{\mathrm{E},N} \vert \lessapprox 0.5$ (see Figure~\ref{fig:pdf_PI_E}). Additionally, fitting the models separately to OGLE and MOA datasets resulted in distinctly different $\pi_{\mathrm{E}}$ posteriors (Figure~\ref{fig:PI_E}).
\subsection{Galactic prior}
For the new OGLE reductions, we repeated additional analyses conducted for the initial reductions (including the addition of the xarallap effect, the 2L2S models, and the Gaussian Processes correlated noise modeling). However, none of these approaches resulted in reliable improvements in $\chi^2$ or in the values of $\pi_{\mathrm{E}}$. For these reasons, we implemented in our modeling priors on  $\pi_{\mathrm{E},N}$  and $\pi_{\mathrm{E},E}$ based on the Galactic model from \cite{2021ApJ...917...78K} and \cite{2022zndo...6869520K}.  We estimated the input event parameters for the Galactic model by summing the distributions of all three 2L1S $u_0+$ topologies ("close", "medium", and "wide"), and considered the $3\sigma$ posterior scatter of the selected parameters, as summarized in Table~\ref{tab:gal_prior}. 
With those settings, we simulated a sample of $5\times 10^7$ microlensing events with the \texttt{genulens} code \citep{2022zndo...6869520K}.  As priors for the Bayesian inference we used the resulting samples  of  $\pi_{\mathrm{E},N}$  and $\pi_{\mathrm{E},E}$, which were broadened in order to obtain proper sampling of wings of the posterior distribution.  We achieved that by averaging simulated samples with normal distributions with  $\mu= 0.05$, $\sigma=0.30$  and $\mu= 0.03$, $\sigma=0.28$, respectively.
These values of $\mu$ and $\sigma$ were determined by fitting Gaussian functions to the simulated $\pi_{\mathrm{E},N}$ and $\pi_{\mathrm{E},E}$ distributions, with the resulting~$\sigma$ values scaled by a factor of 4.    
Figure~\ref{fig:pdf_PI_E} presents the priors alongside the original probability distribution functions of the simulated events. The parameters resulting from the modeling are summarized in Table~\ref{tab:gal}, and the obtained distributions of $\pi_{\rm{E}}$ values are included in Figure~\ref{fig:PI_E}.
\begin{table*}
\caption{Settings of the Galactic model \citep{2021ApJ...917...78K}  used in simulating microlensing events.} 
\label{tab:gal_prior}
\centering 
\begin{tabular}{c|c}
\hline\hline Setting & Explanation \\
\hline 
\makecell{$v_{\oplus,\mathrm{N}} =-1.61~\mathrm{km~s^{-1}} $\\ $v_{\oplus,\mathrm{E}} =  10.56 ~\mathrm{km~s^{-1}}$ }&
\makecell{Earth velocity at $\mathrm{BJD_{TT}}=2457267.5$ projected on \\ the  plane of sky towards event's coordinates.}\\
$22.30~\rm{d} < t_{\rm{E}} < 35.82 ~\rm{d} $ &The Einstein crossing time.\\
$17.04~\rm{mag} < I_{\rm{S}} <  17.84~\rm{mag} $   & Brightness of the source in the OGLE  $I$ -- band filter. See Section 4.2.1. \\
$2.86~\rm{mag}<(V-I)_{\rm{S}}<3.03~\rm{mag}$ & $(I-V)$ color of the source. See Section 4.2.1. \\
$A_{\rm{I,RC}} = 2.08~\rm{mag} $ & Mean Red Clump extinction in the target field.  See Section 4.2.1.\\
$E(V-I)_{\rm{RC}} = 1.88 ~\rm{mag}$ & Mean Red Clump reddening in the target field.  See Section 4.2.1.\\
\hline 
\end{tabular}
\end{table*}
\begin{table*}
\caption{Parameter values with uncertainties of the best 2L1S+parallax model topologies for OGLE-2015-BLG-1609, fitted to the OGLE data alone, using a Galactic model prior on $\pi_{\rm{E}}$.}
\label{tab:gal}
\centering 
\begin{tabular}{c|c|c|c}
\hline\hline 
 & close $u_{\rm{0}}+$ & medium $u_{\rm{0}}+$ & wide $u_{\rm{0}}+$ \\
\hline 
$\chi^{2}/\rm{d.o.f.}$ & 822.2/831 & 832.0/831 & 807.1/831 \\
Prior probability density & 7.10 & 6.97 & 7.93 \\
$q$ & $0.00137^{+0.00032}_{-0.00044}$ & $0.000241^{+0.000024}_{-0.000017}$ & $0.00121^{+0.00033}_{-0.00029}$ \\
$s$ & $1.087^{+0.060}_{-0.030}$ & $1.2310\pm0.0091$ & $1.375^{+0.041}_{-0.036}$ \\
$t_{\rm 0}$ [HJD] & $2457261.882^{+0.116}_{-0.074}$ & $2457261.807^{+0.105}_{-0.076}$ & $2457261.913^{+0.109}_{-0.073}$ \\
$u_{\rm 0}$ & $0.319\pm0.011$ & $0.3255\pm0.0100$ & $0.332\pm0.011$ \\
$t_{\rm E}$ [d]& $28.09\pm0.62$ & $28.01\pm0.60$ & $27.65\pm0.63$ \\
$\rho$ & $0.033^{+0.016}_{-0.020}$ & $0.0479^{+0.0024}_{-0.0018}$ & $0.0376^{+0.0068}_{-0.0132}$ \\
$\alpha$ [deg] & $309.88\pm0.51$ & $309.24^{+0.62}_{-0.52}$ & $308.93\pm0.47$ \\
$\pi_{\mathrm{E},N}$\tablefootmark{a} & $-0.01^{+0.14}_{-0.31}$ & $0.00^{+0.15}_{-0.28}$ & $-0.01^{+0.14}_{-0.29}$ \\
$\pi_{\mathrm{E},E}$\tablefootmark{a} & $-0.022^{+0.048}_{-0.042}$ & $-0.047\pm0.042$ & $-0.035^{+0.048}_{-0.042}$ \\
\hline 
\end{tabular}
\tablefoot{
\tablefoottext{a}{For the reference time $t_{\mathrm{0,par}}=2457267.5$ HJD}}
\end{table*}
\section{Final multi-instrument data modeling -- OGLE, MOA, RoboNet, MiNDSTEp}
\subsection{Microlens parameters}
To conclude the analysis of OGLE-2015-BLG-1609 and obtain trustworthy parameters of the system, we fitted 2L1S models with the parallax effect simultaneously to all available data from the four mentioned projects: OGLE, MOA, RoboNet, MiNDSTEp.
In all three topologies (“close”, “medium”, “wide”), the models with positive values of $u_{\rm{0}}$ had lower values of $\chi^2$ statistics. These three models are plotted in Figure~\ref{fig:model}, and Figure~\ref{fig:modle_zoom} shows the same models zoomed in on the planetary anomaly. The corresponding parameters of these models, along with their $\chi^2$ statistics, are listed in Table~\ref{tab:parms}.
Error bars scaling parameters for corresponding topologies are summarized in Table~\ref{tab:err}.

\begin{table*}
\caption{Parameter values with uncertainties of the best 2L1S+parallax model topologies for OGLE-2015-BLG-1609, fitted to all available datasets.}
\label{tab:parms}
\centering     
\begin{tabular}{c|c|c|c}
\hline\hline 
&close $u_{\rm{0}}+$ & medium $u_{\rm{0}}+$ &wide $u_{\rm{0}}+$\\
\hline  
$\chi^{2}/\rm{d.o.f.}$ & 3251.1/3405 & 3239.4/3405 & 3248.3/3405 \\
$q$ & $0.00123\pm0.00019$ & $0.000225^{+0.000032}_{-0.000026}$ & $0.00133^{+0.00018}_{-0.00016}$ \\
$s$ & $1.069^{+0.023}_{-0.017}$ & $1.2106^{+0.0076}_{-0.0061}$ & $1.416\pm0.013$ \\
$t_{\rm 0}$ [HJD]& $2457261.92^{+0.21}_{-0.26}$ & $2457261.77^{+0.22}_{-0.25}$ & $2457261.98^{+0.20}_{-0.25}$ \\
$u_{\rm 0}$ & $0.3056\pm0.0072$ & $0.3138^{+0.0067}_{-0.0076}$ & $0.3180\pm0.0074$ \\
$t_{\rm E}$ [d] & $29.12^{+0.73}_{-0.51}$ & $29.00^{+0.90}_{-0.57}$ & $28.54^{+0.70}_{-0.49}$ \\
$\rho$ & $0.018^{+0.015}_{-0.013}$ & $0.0469^{+0.0019}_{-0.0025}$ & $0.0085^{+0.0089}_{-0.0060}$ \\
$\alpha$ [deg]& $309.63^{+0.62}_{-0.45}$ & $309.83^{+0.65}_{-0.51}$ & $308.55^{+0.56}_{-0.42}$ \\
$\pi_{\mathrm{E},N}$\tablefootmark{a} & $-0.07^{+0.63}_{-0.55}$ & $0.21\pm0.57$ & $-0.06^{+0.62}_{-0.54}$ \\
$\pi_{\mathrm{E},E}$\tablefootmark{a} & $-0.030^{+0.051}_{-0.038}$ & $-0.059^{+0.043}_{-0.033}$ & $-0.033^{+0.051}_{-0.041}$ \\
\hline
\end{tabular}
\tablefoot{
\tablefoottext{a}{For the reference time $t_{\mathrm{0,par}}=2457267.5$ HJD}}
\end{table*}

\begin{table*}
\caption{Error bars scaling parameters (defined as in Equation \ref{equ:err}) with uncertainties for the best 2L1S+parallax model topologies for the OGLE-2015-BLG-1609 event, fitted to all available datasets 
$e_{\rm{min}}$ values are reported in magnitudes units.}
\label{tab:err}
\centering
\begin{tabular}{c|c|c|c}
\hline\hline                     
&close $u_{\rm{0}}+$  & medium $u_{\rm{0}}+$ & wide $u_{\rm{0}}+$\\
\hline            
$k_{\rm{OGLE~I-band}}$ & $1.269\pm0.100$ & $1.200\pm0.101$ & $1.19\pm0.11$ \\
$e_{\rm{OGLE~I-band}}$ & $0.0043^{+0.0011}_{-0.0015}$ & $0.00510^{+0.00094}_{-0.00113}$ & $0.00537^{+0.00095}_{-0.00108}$ \\
$k_{\rm{OGLE~V-band}}$ & $1.32^{+0.39}_{-0.59}$ & $1.31^{+0.41}_{-0.67}$ & $1.32^{+0.40}_{-0.61}$ \\
$e_{\rm{OGLE~V-band}}$ & $0.0109^{+0.0091}_{-0.0077}$ & $0.0117^{+0.0092}_{-0.0083}$ & $0.0108^{+0.0093}_{-0.0076}$ \\
$k_{\rm{MOA~R-band}}$ & $1.179\pm0.018$ & $1.182\pm0.018$ & $1.183\pm0.019$ \\
$e_{\rm{MOA~R-band}}$ & $0.0016^{+0.0014}_{-0.0011}$ & $0.0016^{+0.0015}_{-0.0011}$ & $0.0017^{+0.0015}_{-0.0012}$ \\
$k_{\rm{Danish~\sim i+z-band}}$ & $6.06^{+0.74}_{-1.40}$ & $6.08^{+0.71}_{-1.36}$ & $6.04^{+0.74}_{-1.41}$ \\
$e_{\rm{Danish~\sim i+z-band}}$ & $0.020^{+0.022}_{-0.014}$ & $0.020^{+0.021}_{-0.014}$ & $0.021^{+0.022}_{-0.015}$ \\
$k_{\rm{R-LCO~I-band}}$ & $4.8^{+2.1}_{-3.2}$ & $4.7^{+2.1}_{-3.1}$ & $5.0^{+2.0}_{-3.2}$ \\
$e_{\rm{R-LCO~I-band}}$ & $0.037^{+0.016}_{-0.024}$ & $0.038^{+0.014}_{-0.024}$ & $0.036^{+0.015}_{-0.024}$ \\
$k_{\rm{T-LCO~I-band}}$ & $1.96\pm0.93$ & $1.93\pm0.92$ & $1.91\pm0.93$ \\
$e_{\rm{T-LCO~I-band}}$ & $0.0140^{+0.0118}_{-0.0097}$ & $0.0140^{+0.0116}_{-0.0094}$ & $0.0142^{+0.0115}_{-0.0097}$ \\
$k_{\rm{S-LCO~I-band}}$ & $6.48^{+1.09}_{-0.91}$ & $6.46^{+1.07}_{-0.94}$ & $6.48^{+1.09}_{-0.92}$ \\
$e_{\rm{S-LCO~I-band}}$ & $0.082^{+0.085}_{-0.056}$ & $0.086^{+0.089}_{-0.060}$ & $0.082^{+0.088}_{-0.057}$ \\
\hline                                       
\end{tabular}
\end{table*}
\subsection{Physical parameters}
\subsubsection{Source star}
\begin{figure}[htb!]
\centering
\includegraphics[width=1\linewidth]{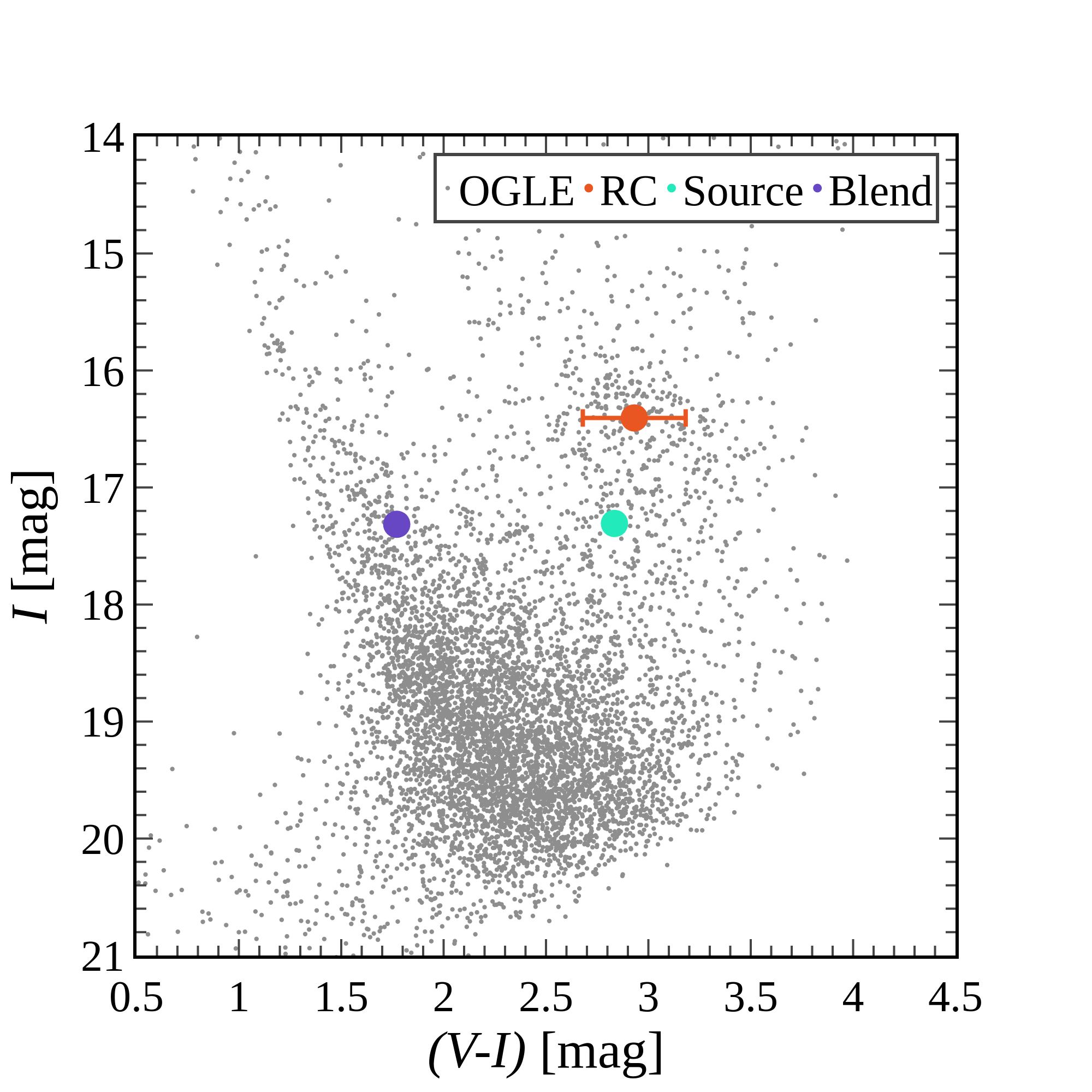}
\caption{Color-magnitude diagram for stars in the OGLE-IV data within $2\arcmin$ around the microlensed star in the OGLE-2015-BLG-1609 event. The red circle marks the position of the RC centroid, while the turquoise and purple circles mark the positions of the source and blend, respectively.}
\label{fig:CMB}  
\end{figure}

Based on the results from the microlensing modeling, the source star is most likely a typical Galactic bulge red giant. In contrast, the blend is a bluer main-sequence star (see Figure~\ref{fig:CMB}).\par
To acquire extinction parameters, we compared the position of the Red Clump (RC) Giants centroid on the CMD to its intrinsic color and brightness. Based on the OGLE-IV photometric maps for the Galactic bulge, calibrated to the standard systems, we selected stars within  2\arcmin~around the location of the event and lying on the Red Giant Branch, i.e., $(V-I)> 2.4$ mag and $16.1< I< 16.7$ mag (Figure~\ref{fig:CMB}). This criterion provided a sample of $N_{\rm{stars}}=543$. Using these stars, we fitted the luminosity function parameterized as in \cite{2013ApJ...769...88N}:
\begin{equation}
\begin{aligned}
N(I)dI &= A\exp\biggl[B(I-I_{\rm{RC}})\biggl]  +\\ &
 \frac{N_{\rm{RC}}}{\sqrt{2\pi}\sigma_{\rm{RC}}}\exp \biggl[{-\frac{(I-I_{\rm{RC}})^2}{2\sigma_{\rm{RC}}^2}}\biggl] 
+ \\ & 
 \frac{N_{\rm{RGBB}}}{\sqrt{2\pi}\sigma_{\rm{RGBB}}}\exp \biggl[{-\frac{(I-I_{\rm{RGBB}})^2}{2\sigma_{\rm{RGBB}}^2}}\biggl]  
+ \\ & \frac{N_{\rm{AGBB}}}{\sqrt{2\pi}\sigma_{\rm{AGBB}}}\exp \biggl[{-\frac{(I-I_{\rm{AGBB}})^2}{2\sigma_{\rm{AGBB}}^2}}\biggl],
\end{aligned}
\label{EQ:Exponential}
\end{equation}
\begin{eqnarray}
N_{\rm{RGBB}} = 0.201\times N_{\rm{RC}},  \\
N_{\rm{AGBB}} = 0.028\times N_{\rm{RC}},  \\
I_{\rm{RGBB}} = I_{\rm{RC}}+0.737, \\
I_{\rm{AGBB}} = I_{\rm{RC}}-1.07,  \\
\sigma_{\rm{RGBB}} = \sigma_{\rm{AGBB}} = \sigma_{\rm{RC}},
\label{EQ:bumparameters}
\end{eqnarray}
where $I$, $\sigma$, and $N$ are the mean magnitude, magnitude dispersion, and
number of stars, respectively. The subscripts AGBB denote the asymptotic
giant branch bump and RGBB represents the red giant branch bump. Following \cite{2013ApJ...769...88N, 2016MNRAS.456.2692N}, we imposed priors summarized in Table~\ref{tab:RC_prior}. \par
\begin{table}
\caption{Priors used for the RC centroid fitting.}\label{tab:RC_prior}
\centering 
\begin{tabular}{c|c}
\hline\hline 
Parameter&Prior\\
\hline 
$B$&$\mathcal{N}(0.55,0.03)$\\
$N_{\rm{RC}}/A$&$\mathcal{N}(1.17,0.07)$\\
$I_{\rm{RC}}$ & $\mathcal{U}(16,18)$\\
$\int N(I)dI = N_{\rm{exp}} $ & $\mathcal{N}(N_{\rm{stars}},0.5)$ \\
\hline                                         
\end{tabular}
\end{table}
To sample the posterior distribution, we used the \texttt{emcee} code. Assuming a Poisson distribution of observed stars, we employed the likelihood function in the form:
\begin{equation}
    \ln{ \mathcal{L}}  \propto  - N_{\rm{exp}} + \sum_i \ln{N(I_i})dI.
\end{equation}
The fitting resulted in $I_{\rm{RC}}=16.405\pm0.47$ mag and $(I-V)_{\rm{RC}}=2.93\pm0.25$ mag. As seen in the CMD (Figure~\ref{fig:CMB}), interstellar extinction within the selected 2\arcmin~ radius is highly differential, which is typical for the Galactic bulge region \citep{2013ApJ...769...88N}.
We estimated the reddening parameters using the values of intrinsic apparent magnitude of the RC calculated by \cite{2016MNRAS.456.2692N}:
\begin{eqnarray}
I_{\rm{RC},0} = 14.3955 - 0.0239 ~l + 0.0122 ~|b|=14.3350~\rm{mag} \quad\mathrm{and}\\(V-I)_{\rm{RC},0}=1.06~\rm{mag},
\end{eqnarray} and found:
\begin{equation}
    A_{\mathrm{I}}=2.07\pm 0.47 ~\mathrm{mag} \quad \mathrm{and} \quad E(V-I)= 1.87\pm 0.25~\rm{mag}.
\end{equation}\par
We combined the reddening with the posterior distribution of the source flux from the microlensing modeling. From this, we obtained the dereddened color and brightness of the source star. For the "wide $u_{\rm{0}}+$" topology we found:
\begin{equation}
    I_{\mathrm{0}}=15.211\pm0.058~\mathrm{mag} \quad\mathrm{and}\quad (V-I)_{\mathrm{0}}=0.96\pm 0.25~\mathrm{mag}.
    \label{equ:intr}
\end{equation}
\subsubsection{Source Radius}
As the color--angular size relations are typically better constrained for the $K$ -band filter, we transformed $(V-I)_0$ of the source to $(V-K)_0$ using color-color relations from \cite{1988PASP..100.1134B}. To calculate the source angular radius corrected for the limb-darkening, we used relations from \cite{2018MNRAS.473.3608A}:
\begin{equation}
    \log \theta_{*,\rm{LD}} =  0.562\pm0.009 + 0.051\pm0.003~(V - K)_0 - 0.2~K_0 -\log{2}.
    \label{equ:theta}
\end{equation} 
We assumed that in this relation, as well as in the following ones (\ref{equ:teff}, \ref{equ:logg}), the posterior distribution of a given factor is approximated as a normal distribution. 
Taking the calculated distribution of  $I_0$ and $(V-I)_0$ for the "wide $u_{\rm{0}}+$" model provided:
\begin{equation}
    \theta_{*,\rm{LD}}=3.84^{+0.85}_{-0.56}~\mathrm{\mu as}.
\end{equation}
\subsubsection{Limb-darkening Coefficients}
To find the effective temperature of the source star, we used the empirical color relations from \cite{2000AJ....119.1448H} for cold giant stars:
\begin{equation}
\begin{aligned}
T_{\rm{eff}} &= 8556.22\pm204.27-5235.57\pm352.83~(V-I)_0 +\\&1471.09\pm148.20~(V-I)_0^2= 4957_{-720}^{+744} ~\rm{K}.
\end{aligned}
\label{equ:teff}
\end{equation}
Estimation of surface gravity was obtained from \cite{1994AstL...20..755B} the relation for G-K giants and subgiants, assuming solar metallicity of the source:
\begin{equation} 
\begin{aligned}
    \log{g} &= 8.0\pm0.04~T_{\rm{eff}} + 0.31\pm0.04~ M_{\rm{0,V}} +\\&0.27\pm0.11~\rm[Fe/H]-27.15\pm2.1 =3.1\pm 2.2,
\end{aligned}
\label{equ:logg}
\end{equation}
where we used $M_{0,V}= 1.57 \pm 0.27 $ mag.
To derive the limb-darkening coefficients, we used tables from \cite{2011A&A...529A..75C} and assumed solar metallicity and turbulent velocity of $2~\rm{km~s^{-1}}$. Those parameters have minor impact on the values of coefficients. For bands in which observational data were collected, we obtained:
\begin{equation}
    u_V=0.9198 ,\quad u_R= 0.6535  ,\quad u_I=0.5597	,\quad u_{z'}=0.5131.
\end{equation}
\subsubsection{Lens Mass and Distance}
We obtained the angular Einstein radius based on the scaled  source radius from finite-source effect modeling and calculated the observed source radius:
\begin{equation}
    \theta_{\rm{E}}= \frac{ \theta_{*,\rm{LD}}}{\rho}=0.46^{+1.06}_{-0.24} ~\mathrm{mas}.
\end{equation}
Combining this with constraints on the microlensing parallax led to the mass measurement of the lens and the companion \citep{2000ApJ...542..785G}:
\begin{equation}
M_{\rm{L}}=\frac{\theta_{\rm{E}}}{\kappa\pi_{\mathrm{E}}}=0.17^{+0.63}_{-0.12}~  \mathrm{M_{\odot}} \quad\mathrm{and}\quad  M_{\rm{C}}=qM_{\rm{L}} =0.24^{+0.90}_{-0.17} ~\mathrm{M_{\rm{J}}},
\end{equation}
where $\kappa \equiv 4 G/(c^{2} \mathrm{au}) \simeq 8.144~\rm{mas~M}_\odot^{-1}$. Additionally, the source-lens relative parallax is given by:
\begin{equation}
 \pi_{\mathrm{rel}}=\pi_{\rm{E}}  \theta_{\rm{E}}=0.18^{+0.51}_{-0.13} ~\mathrm{mas},
\end{equation}
and the relative proper motion by:
\begin{equation}
    \mu_{\rm{rel}}= \frac{ \theta_{\rm{E}}}{ t_{\rm{E}}}= 5.7^{+13.1}_{-3.0} ~\mathrm{mas~yr^{-1}}.
\end{equation}
Assuming that the source star is at the typical distance for a Galactic bulge star $D_{\rm{S}}=8.54$ kpc, we estimated the lens distance as:
\begin{equation}
    D_{\rm{L}}=\frac{\mathrm{au}}{ \pi_{\rm{rel}}+\pi_{\rm{S}}}= 3.3^{+2.6}_{-2.1} ~\mathrm{kpc},
\end{equation}
where $\pi_{\rm{S}}=1/D_{\rm{S}}$. All the values in this section were calculated for the "wide $u_{\rm{0}}+$" topology. Results for all the considered models are summarized in Table~\ref{tab:parms_lens}. 
\begin{table*}
\caption{\label{tab:parms_lens}Physical properties of OGLE-2015-BLG-1609, including the estimated evidence and the probability for the lens being a substellar object.}
\centering                                     
\begin{tabular}{c|c|c|c|c}
\hline\hline
Data&\multicolumn{3}{c|}{Multi-instrument}&OGLE+Galactic prior\\
\hline
&close $u_{\rm{0}}+$&medium $u_{\rm{0}}+$ &wide $u_{\rm{0}}+$&wide $u_{\rm{0}}+$\\
\hline                       
$\theta_{\rm{E}}$ [mas] & $0.212^{+0.460}_{-0.098}$ & $0.082^{+0.018}_{-0.012}$ & $0.46^{+1.06}_{-0.24}$ & $0.111^{+0.062}_{-0.026}$\\
$\mu_{\rm{rel}}$ [mas/yr] & $2.6^{+5.6}_{-1.2}$ & $1.00^{+0.22}_{-0.15}$ & $5.7^{+13.1}_{-3.0}$& $1.43^{+0.79}_{-0.33}$  \\
$M_{\rm{h}} \rm{[M_{\odot}]}$ & $0.079^{+0.279}_{-0.052}$ & $0.025^{+0.050}_{-0.012}$ & $0.17^{+0.63}_{-0.12}$ & $0.118^{+0.188}_{-0.079}$\\
$M_{\rm{c}} \rm{[M_{J}]}$ & $0.103^{+0.384}_{-0.070}$ & $0.0059^{+0.0122}_{-0.0031}$ & $0.24^{+0.90}_{-0.17}$ & $0.140^{+0.273}_{-0.095}$  \\
$D_{\rm{L}}$ [kpc] & $4.9^{+2.1}_{-2.6}$ & $6.6^{+1.2}_{-1.3}$ & $3.3^{+2.6}_{-2.1}$ & $7.51^{+0.60}_{-1.53}$\\
$\log{\mathcal{Z}}$&6.46 &5.63 & 6.51 &6.16\\
$ Pr(M_{\mathrm{L}}<80 \mathrm{M_{J}})$ &  $49\%$ & $85\%$ & $25\%$& $35\%$\\
\hline  
\end{tabular}
\end{table*}
\subsection{Comparison of solutions}
Table~\ref{tab:parms} includes the $\chi^2$ values for all three topologies. There is only a minor difference in $\chi^2$ between the "close" and "wide" topologies. The "medium" topology has a slightly lower $\chi^2$ ($\Delta\chi^2 \approx 10$). This difference in $\chi^2$ is traced to observations from a single night (HJD = 2457267; Figure~\ref{fig:modle_zoom}). On that night, the rising slope of the planetary anomaly in the "medium" model fits all the OGLE data points, unlike the other two topologies and unlike other nights, where similar variations occurred between measurements. Specifically, in the case of the "medium" topology, the four OGLE observations taken that night contribute $\chi^2/\mathrm{d.o.f.}= 0.9/4$ to the overall $\chi^2/\mathrm{d.o.f.}=  3239.4/3405$. The OGLE observing logs indicate no anomalous conditions during that night, although such a perfect model fit to the data points seems unlikely and indicates overfitting of the "medium" topology. Unfortunately, no observations from other projects were collected that night to verify the shape of the light curve seen in the OGLE data.\par
To quantitatively verify the model topologies, we estimated the Bayesian evidence for each of them. This was done using previously simulated microlensing events from the Galactic model of \cite{2021ApJ...917...78K} with event input properties given in Table~\ref{tab:gal_prior}.  Additionally, from the posterior distribution of $\theta_{\rm{E}}$ for each topologies, we derived the corresponding probability density functions.
The evidence was then calculated as:
\begin{equation}
    \mathcal{Z}= \sum_i p_i w_i \quad \mathrm{with} \quad  p_i=f(\theta_{\mathrm{E},i} | \mathrm{model}), 
\end{equation}
where index $i$ denotes $i-$th simulated event, $w_i$ is its weight, and $f$ is derived probability density function for $\theta_{\rm{E}}$ for a given model. Evidence values are listed in Table~\ref{tab:parms_lens}. Based on these estimates, only the "medium" topologies can be considered as significantly less probable, while the differences between the "close" and "wide" topologies remains inconclusive.\par
Considering that $80~\mathrm{M_{J}}$ is the mass limit for sustaining hydrogen fusion into helium in a stellar core, all topologies suggest that the lens could be a substellar, brown dwarf object. For the most likely "close" and "wide" topologies, this probability is estimated to be $49\%$ and $25 \%$, respectively, as shown in Table~\ref{tab:parms_lens}. Using Bayesian evidence values as proportionality indicators, we estimated that the overall probability for lens to be under the $80~\mathrm{M_{J}}$ mass limit is $34\%$, in case of the $u_{0}+$ models.
Taking into account the three major planetary detection methods: transit, radial velocity, and microlensing; OGLE-2015-BLG-1609L is the sixteenth known system whose host is suspected to fall within the brown dwarf mass range \citep{refId0}. All other systems were also detected through microlensing, which results from the unique capabilities of this method in detecting faint objects. Furthermore, the low mass of the lens suggests that the blended flux cannot be fully explained by the lens's light contribution.\par
To better characterize the lens system and more accurately constrain the lens mass, additional high-angular-resolution observations will be needed. Given the relative proper motion of approximately $\mu_{\mathrm{rel}}=5~\mathrm{mas~yr^{-1}}$ and the angular resolution of dedicated adaptive optics instruments of $60~\mathrm{mas}$ \citep{2023arXiv230201168V}, we could potentially detect the lens by performing observations in 2027 or later. If the lens's light remains undetected, this would further suggest its substellar nature. 
This approach was used for the MOA-2010-BLG-447 event, where the non-detection of lens flux using near-infrared observations from the Keck Observatory strongly supported a white dwarf interpretation for the lens \citep{2021Natur.598..272B}. 
Alternatively, use of dual-field interferometer instrument GRAVITY to detect the blend flux, attribute it to the lens, and determine the microlensing system parameters is not possible because of a lack of a bright star ($K \lesssim 10$ mag) for fringe tracking within $30 \arcsec$ of the event's sky location.
\section{Summary}
We analyzed the microlensing event OGLE-2015-BLG-1609, with a planetary anomaly prominent in the surveys data alone. We used a series of modeling approaches, including imposing Galactic model priors, to mitigate challenges such as the unconstrained parallax in the survey data. After incorporating data from the MOA and follow-up projects, and refining the OGLE reductions, we identified three possible model topologies for the microlensing system. Of these, two more likely ones have indistinguishable Bayesian evidence. Ultimately, our analysis points to a  probability of 34\% that the lens system is a planet-hosting brown dwarf. We conclude that with a relative proper motion on the order of $\mu_{\mathrm{rel}}=5~\mathrm{mas~yr^{-1}}$, if the lens's flux is not resolved in the near future by high-angular-resolution instruments, the probability of the lens being a substellar object will increase.

\begin{acknowledgements}
OGLE Team acknowledges  former members of the team, for their contribution to the collection of the OGLE photometric data over the past years. The work by Radosław Poleski was partly supported by the Polish National Agency for Academic Exchange grant “Polish Returns 2019”.The MOA project is supported by JSPS KAKENHI Grant Number 
JP24253004, JP26247023,JP16H06287 and JP22H00153.
YT acknowledges the support of the DFG priority program SPP 1992 "Exploring the Diversity of Extrasolar Planets (TS 356/3-1)
R.F.J. acknowledges support for this project provided by ANID's Millennium Science Initiative through grant ICN12\textunderscore 009, awarded to the Millennium Institute of Astrophysics (MAS), and by ANID's Basal project FB210003. L.M. acknowledges financial contribution from PRIN MUR 2022 project 2022J4H55R.

\end{acknowledgements}

%
\bibliographystyle{aa} 
\bibliography{bib} 
%

%

\end{document}